# Temperature-dependent elastic constants of thorium dioxide probed using time-domain Brillouin scattering

Amey Khanolkar[1,*], Yuzhou Wang[1], Cody A. Dennett[1], Zilong Hua[1], J. Matthew Mann[2], Marat Khafizov[3], and David H. Hurley[1,§]

[1]Condensed Matter and Materials Physics Group, Idaho National Laboratory, Idaho Falls ID 83415
[2]Air Force Research Laboratory, Sensors Directorate, Wright-Patterson Air Force Base OH 45433
[3]Department of Mechanical and Aerospace Engineering, The Ohio State University, Columbus OH 43210

Corresponding authors: [*]Amey.Khanolkar@inl.gov (A. Khanolkar), [§]David.Hurley@inl.gov (D.H. Hurley)

## ABSTRACT

We report the adiabatic elastic constants of single-crystal thorium dioxide over a temperature range of 77 – 350 K. Time-domain Brillouin scattering (TDBS), an all-optical, non-contact picosecond ultrasonic technique, is used to generate and detect coherent acoustic phonons that propagate in the bulk perpendicular to the surface of the crystal. These coherent acoustic lattice vibrations have been monitored in two hydrothermally grown single-crystal thorium dioxide samples along the (100) and (311) crystallographic directions. The three independent elastic constants of the cubic crystal ($C_{11}$, $C_{12}$ and $C_{44}$) are determined from the measured bulk acoustic velocities. The longitudinal wave along the (100) orientation provided a direct measurement of $C_{11}$. Measurement of $C_{44}$ and $C_{12}$ was achieved by enhancing the intensity of quasi-shear mode in a (311) oriented crystal by adjusting the polarization angle relative to the crystal axes. We find the magnitude of softening of the three elastic constants to be ~2.5% over the measured temperature range. Good agreement is found between the measured elastic constants with previously reported values at room temperature, and between the measured temperature-dependent bulk modulus with calculated values. We find that semi-empirical models capturing lattice anharmonicity adequately reproduce the observed trend. We also determine the acoustic Gruneisen anharmonicity parameter from the experimentally derived temperature-dependent bulk modulus and previously reported temperature-dependent values of volume thermal expansion coefficient and heat capacity. This work presents measurements of the temperature-dependent elasticity in single-crystal thorium dioxide at cryogenic temperature and provides a basis for testing *ab initio* theoretical models and evaluating the impact of anharmonicity on thermophysical properties.

## I. INTRODUCTION

Physical properties of crystals are governed by interatomic interactions within the lattice, which have a potential energy that is determined by the spatial positions of the nuclei [1]. The interatomic potential energy is typically expanded as a power series of the displacements of the nuclei from their equilibrium positions. As the temperature of a crystalline solid is increased, the average kinetic energy of the atoms in the lattice rises, as does their displacement from their ideal lattice positions. Several characteristics of real crystals cannot be explained using the harmonic approximation that assumes small atomic displacements described only by the quadratic terms of the potential energy. For instance, within the harmonic approximation, there would be no thermal expansion, the heat capacity would be temperature-independent at high temperatures, and heat transport would be ballistic. Including the effects of lattice anharmonicity, represented by the higher order terms in the interatomic potential, is therefore central to accurately modelling vibrational properties of a variety of crystalline materials [2, 3].

Thorium dioxide (ThO$_2$), also known as thoria, is a crystalline compound that has garnered considerable attention as an alternative nuclear fuel to traditional urania (UO$_2$) owing to its lower thermal expansion coefficient and higher thermal conductivity, melting point and oxidation resistance, all of which are essential fuel properties for enhanced reactor safety [4, 5]. In contrast to urania, however, there exist very limited experimental data on the thermophysical and thermodynamic properties of thoria [6]. Such data are essential for developing a fundamental understanding of the performance of thorium-bearing fuels in next-generation nuclear reactors [6]. Several studies have used first-principles calculations and atomistic simulations to predict elastic, thermophysical and thermodynamic properties of thoria [7-20]. Recently, an irreducible derivatives approach to parameterize the quash-harmonic approximation for interacting phonons was implemented, and the temperature-dependent elastic constants and thermal expansion coefficient of thoria were computed using density functional theory (DFT) with three exchange-correlation functionals [21]. Measurement of the temperature- or pressure-dependence of fundamental material properties can provide stringent tests for these theoretical calculations involving anharmonic phonon interactions [3].

The elastic constants, $C_{ij}$, are a fundamental property of a crystalline solid that are represented by components of a fourth-rank tensor and defined as the second spatial derivative of the interatomic potential. Anharmonicity in the interatomic potential function causes thermal expansion, and as a result, the elastic constants exhibit temperature-dependence due to the strain induced by thermal expansion. Measuring the temperature-dependence of the elastic constants is therefore a convenient approach to probe anharmonicity in the crystal lattice. Leibfried and Ludwig [1] derived an expression for the temperature-dependence of the adiabatic elastic constants:

$$C_{ij}(T) = \widetilde{C}_{ij}(1 - D_{ij}\bar{\epsilon}) \qquad (i)$$

Here, $\widetilde{C}_{ij}$ represents the zero-temperature elastic constants in the harmonic approximation, $\bar{\epsilon}$ is the mean thermal energy of an oscillator at temperature $T$, and $D_{ij}$ is a constant and captures anharmonicity that depends on the type of crystal and model used. Numerous studies have experimentally investigated the temperature-dependence of elastic constants in single crystals. Early studies employed the ultrasonic pulse-echo technique to measure the elastic constants as a function of temperature in materials such as GaAs [22], GaSb and GaP [23], sodium [24], rubidium [25], yttrium [26], and cadmium fluoride [27] single crystals, as well as polycrystalline indium [28]. Others have used resonant ultrasound spectroscopy (RUS) to evaluate elastic constants from the measured free-body resonant frequencies of the sample at various temperatures [29-33]. Sonehara et al. used stimulated Brillouin spectroscopy to measure the temperature dependence of the Brillouin frequency acoustic wave frequency shift in TeO$_2$ and PbMoO$_4$ crystals [34]. The temperature-dependence of the elastic constants of taurine (2-aminoethanesulfonic acid) single crystals was recently measured by Kang et al. using a Brillouin spectrometer equipped with a tandem Fabry-Perot interferometer [35]. In most cases, the measured elastic constants as a function of temperature were fitted to semi-empirical models derived from the Leibfried-Ludwig relation, such as those proposed by Varshni [36] (that used the energy of an Einstein oscillator) and by Lakkad [37] (that was based on the Debye model). The Debye temperature of the crystal $\theta_D$, a parameter associated with lattice vibrations that correlates elastic properties with thermodynamic properties such as thermal conductivity, specific heat, thermal expansion, and enthalpy, was also calculated from the measured elastic constants [38] in many of these studies. Several studies also used the elastic constants to determine the acoustic Gruneisen parameter $\gamma$, a dimensionless quantity that represents the dependence of the acoustic phonon frequencies on the system volume, and therefore, the system's anharmonicity.

Thoria has a fluorite crystal structure (space group $Fm\bar{3}m$), in which the thorium cation sublattice forms a face-centered-cubic (FCC) arrangement, while the oxygen anions occupy all the tetrahedral sites. Due to the symmetry of the cubic crystal, the elastic stiffness tensor of thoria has three independent elements ($C_{11}$, $C_{12}$, and $C_{44}$). As stated previously, while several prior studies [7-21] have utilized *ab initio* calculations or atomistic simulations to predict the thermo-physical properties of thoria, there have been very limited experimental measurements of material properties to validate these models, particularly for elastic constants of single crystal thoria. There have been only two studies that have measured the elastic constants of single crystal thoria at room temperature – one by Macedo et al. that used the ultrasonic pulse-echo technique [39], and the other by Clausen et al. that utilized inelastic neutron scattering (INS) to derive the elastic constants from the measured phonon dispersion curves [40]. There is a discrepancy in the values of the room temperature elastic constants reported by these two studies – Macedo et al. measured $C_{11}$ = 367 GPa, $C_{12}$ = 106 GPa and $C_{44}$ = 79 GPa using ultrasonic pulse-echo, while Clausen extracted higher values of $C_{11}$ = 377 GPa, $C_{12}$ = 146 GPa and $C_{44}$ = 89 GPa from INS. While previous studies have investigated the temperature-dependence of the elastic moduli of polycrystalline [41, 42] and porous [43, 44] thoria samples, very limited experimental data are available for measurements on high-quality single crystal thoria. The temperature dependence of the $C_{44}$ elastic stiffness tensor component was recently reported from 300 K to 1200 K using time-of-flight INS [21]. The same study also reported measurements of the temperature dependence of $C_{11}$ from 77 K to 350 K, and the values of $C_{12}$ and $C_{44}$ on single crystal thoria at 77 K. A systematic experimental investigation of the temperature dependence of the three independent elastic constants of thoria at cryogenic temperatures, however, has been lacking.

Here, we report the temperature-dependent elastic constants of single crystal bulk thoria samples, for temperature ranging from 77 K to 350 K, that have been synthesized using the hydrothermal growth technique. An all-optical, non-contact pump-probe picosecond ultrasonic technique, known as time-domain Brillouin scattering (TDBS) [45], is used to generate coherent acoustic phonons that propagate in the bulk of the crystal. Lattice vibrations associated with the acoustic wave are detected optically by a time-delayed probe beam. Unlike the pulse-echo and the RUS techniques, TDBS does not require precise measurements of the sample dimensions, nor does it need a specific sample geometry or clamping boundary conditions. Direct measurement of the ultrasonic wave frequency in TDBS helps overcome uncertainties associated with determining arrival times in the pulse-echo method. The all-optical nature of generation and detection of ultrasound also bypasses the use of contact-based transducers [46]. Compared to INS that maps phonon dispersion curves up to the Brillouin zone edge, the TDBS technique directly probes lattice vibrations associated with long-wavelength acoustic phonons. The elastic constants are calculated from the measured ultrasonic wave speeds. The resultant temperature-dependent bulk modulus is compared to predictions made by computational models [8, 11, 17, 21]. We calculate the Debye temperature from the measured elastic constants and compare it with values obtained from other experimental techniques. We also calculate the temperature-dependent Grueisen parameter [47] from the measured bulk moduli. Finally, we discuss approaches for increased sensitivity to the three elastic constants using TDBS, for an all-encompassing experimental methodology to validate *ab initio* computation methods.

## II. METHODS

*2.1. Sample preparation*

Thoria single crystals were grown using the hydrothermal synthesis technique following previously published procedures [48]. Crystals with large (100) and (311) surface orientations were identified via the crystal morphology and the angle between faces and were mounted parallel with the top surface of copper sample holder using silver paste for optimal thermal contact between the thoria crystal and the sample holder. An additional step of applying Devcon 5-minute epoxy® over the silver paste and along the edges of the crystal prevented a disorientation of the samples during the silver paste curing process. The (100) $ThO_2$ crystal had a mass of 18 mg, while the (311) $ThO_2$ crystal was smaller, with a mass of 6 mg.

*2.2. Time-domain Brillouin Spectroscopy*

Measurements of the crystal elasticity were performed using TDBS [45, 49], wherein coherent acoustic phonons are generated and detected. A schematic representation of phonon generation and detection is illustrated in Figure 1(a). Picosecond duration acoustic pulses propagating normal to the surface were generated by irradiating the sample with ultrafast laser pump pulses (optical wavelength of 780 nm, pulse duration of ~1ps, and pulse repetition rate of 80 MHz) emitted by a titanium-sapphire laser (Coherent Chameleon) and focused to a spot size of ~10 μm using a 20x objective lens [50]. Rapid thermal expansion of the gold film launches coherent acoustic phonons that propagate in the bulk of the thoria crystal along the surface normal. The amplitude of the pump pulse train was modulated at 500 kHz using an acousto-optic modulator (Gooch & Housego). The average power of the pump pulse train was ~10 mW. A thin gold film (~7 nm in thickness) was deposited on the sample surface using DC magnetron sputtering to ensure strong optical absorption of the pump for broadband ultrasonic wave generation by thermal expansion and enhanced shear wave generation through mode conversion [51]. The steady-state local temperature rise due to heating from the pump beam was estimated to be less than 1°C [52]. A time-delayed probe pulse (derived from the same laser source and frequency doubled to an optical wavelength of 390 nm using a beta-barium borate crystal, with average optical power of ~5 mW) was used to detect the ultrasonic pulses propagating in the sample by focusing onto the sample surface through the same optics as the pump beam. Due to optical transparency of thoria at the probe beam wavelength, optical interference between the partial reflection off the free surface of the sample and that from the propagating acoustic pulse (due to photo-elastic coupling) was observed through time-resolved reflectivity changes of the probe beam, referred to as Brillouin oscillations, using a silicon photodiode (Thorlabs PDA55) with a DC to 10 MHz bandwidth. Changes in optical reflectivity of the probe beam) at the pump beam modulation frequency were measured using lock-in detection (Stanford Research Instruments SRS844. The relative temporal delay between the pump and probe pulses was controlled using a mechanical delay line (Newport) placed in the optical path of the probe beam. The temperature dependence of the Brillouin oscillations was measured by placing the sample in a liquid nitrogen-cooled cryostat (Cryo Industries of America) with a transparent window that allowed normal incidence of the pump and probe beams on the sample. The temperature was monitored using an embedded silicon diode temperature sensor installed on the sample holder near the sample. A closed-loop-feedback temperature controller maintained the temperature to within ± 3 K of the setpoint.

The ultrasonic velocity $v$ of the bulk acoustic mode is related to the measured Brillouin oscillation frequency $f$ of that mode, using the equation applicable to normal incidence of the probe beam on the sample surface,

$$v = \frac{f\lambda}{2n} \tag{ii}$$

where $\lambda = 390$ nm is the optical wavelength of the probe beam and $n$ is the real part of the refractive index of thoria at 390 nm [50]. To determine the refractive index $n$ of thoria, we used a spectroscopic ellipsometer (M-2000, J.A. Woollam Co.) with a spectral range of 370 – 1000 nm, and spectral resolution of 1 nm. Reflectance data were acquired at room temperature at seven angles of incidence ($\Phi = 45°, 50°, 55°, 60°, 65°, 70°$, and $75°$) on an uncoated thoria crystal with surface normal along the (100) orientation. The real and imaginary components of the raw reflectance data were fitted to Cauchy and Sellmeier dispersion models [53] to obtain the refractive index of the thoria crystal.

III. RESULTS

*3.1. Generation and detection of coherent acoustic phonons*

Figure 1(b) shows a typical transient reflectivity signal of ~1.5 ns duration measured on a (311) thoria sample. The impulse at t = 0 is due to ultrafast electron heating after absorption of the pump pulse, followed by a gradual decay due to thermal diffusion and electronic relaxation. Gigahertz frequency Brillouin oscillations are superimposed over this thermal/ electronic decay. A moving average-based smoothing function, denoted by the solid red line in Fig. 1(b), was used to fit and subtract the thermal/ electronic background to isolate the Brillouin oscillations. The inset in Fig. 1(b) shows the Brillouin oscillations after subtraction of the decay background. A Fourier transform of the reflectivity transient in Fig. 1(b) is shown in Fig. 1(c). The Fourier amplitude spectrum shows two distinct peaks – a dominant peak at ~64.8 GHz, with another peak at ~34.8 GHz, whose amplitude is about 34% of that of the dominant peak. In the case of the (100) thoria sample, the Brillouin oscillations revealed only a single peak at ~67.3 GHz, corresponding to the longitudinal acoustic mode, as this configuration doesn't allow excitation of the shear modes [50, 51], also shown in Fig. 1(c).

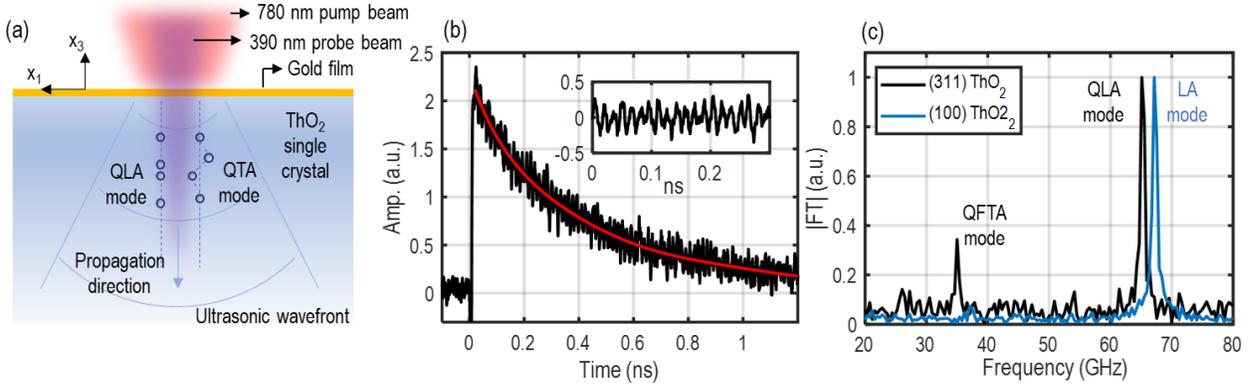

Figure 1. (a) Schematic representation of the all-optical coherent acoustic phonon generation (quasi-longitudinal (QLA) and quasi-transverse (QTA) acoustic modes) in the single crystal thoria sample. (b) Representative time-domain trace showing Brillouin oscillations (black curve) superimposed over the electronic/thermal background (red curve). The inset highlights the Brillouin oscillations after subtraction of the thermal/electronic decay. (c) Corresponding frequency spectrum of the signal in (b) after subtracting the electronic/ thermal background highlighting quasi-longitudinal acoustic (QLA) and quasi-fast transverse acoustic (QFTA) modes in the (311) thoria sample.

*3.2. Velocity and polarization of the measured coherent acoustic phonons*

The ultrasonic velocity of an acoustic mode along any arbitrary direction defined by unit vector $(h_i, h_j, h_k)$ can be obtained from the Christoffel equation by solving the eigenvalue problem [54],

$$\begin{vmatrix} h_1^2 C_{11} + (h_2^2 + h_3^2)C_{44} - \rho v^2 & h_1 h_2 (C_{12} + C_{44}) & h_1 h_3 (C_{12} + C_{44}) \\ h_1 h_2 (C_{12} + C_{44}) & h_2^2 C_{11} + (h_1^2 + h_3^2)C_{44} - \rho v^2 & h_2 h_3 (C_{12} + C_{44}) \\ h_1 h_3 (C_{12} + C_{44}) & h_2 h_3 (C_{12} + C_{44}) & h_3^2 C_{11} + (h_1^2 + h_2^2)C_{44} - \rho v^2 \end{vmatrix} = 0 \quad \text{(iii)}$$

where ρ is the density of the crystal. Using elastic constants $C_{11}$ = 367 GPa, $C_{12}$ = 106 Gpa and $C_{44}$ = 79 Gpa reported by Macedo et al. [39], we calculate the velocity of the longitudinal acoustic (LA) mode along the (100) direction $v_{LA}$ = 6060 m/s. The velocity of the quasi-longitudinal acoustic (QLA) mode (i.e., a mode with primarily longitudinal polarization) along the (311) direction was calculated to be $v_{QLA}$ = 5820 m/s. A fast quasi-transverse acoustic mode (QFTA) along the (311) direction with velocity $v_{QFTA}$ = 3140 m/s is also predicted, as well as a slow quasi-transverse acoustic mode (QSTA) with velocity $v_{QSTA}$ = 2970 m/s. The frequency $f$ of these modes can then be calculated using Eq. (ii). Using spectroscopic ellipsometry, the room-temperature refractive index of thoria at 390 nm was measured to be $n$ = 2.18 ± 0.01. Using the refractive index of thoria, and the ultrasonic velocity obtained from Eq. (iii), we find that the frequency of the LA mode along the (100) orientation is 67.8 GHz. Similarly, the frequencies of the QLA, QFTA and QSTA modes along the (311) direction are 65.1 GHz, 35.1 GHz, and 33.3 GHz, respectively. The frequencies of the acoustic modes detected in the experiment are very close to those calculated from the Christoffel equation using the elastic constants reported by Macedo et al [39]. It is also worth noting that while the Christoffel equation yields two transverse modes QFTA and QSTA that are separated by less than 2 GHz, we only detected one of these two transverse modes along the (311) orientation (discussed in the following section).

*3.3. Optimization of the shear acoustic mode detection amplitude*

Using the opto-acoustic model described in reference [55] that considers the modulation of the dielectric permittivity of the crystal due to the photo-elastic effect, the strain amplitudes of the LA, QFTA and QSTA modes in thoria were calculated along various crystallographic orientations. Inverse pole diagrams that depict the calculated ratios of the strain amplitude of the generated QSTA wave to that of the QLA wave, as well as the ratios of the strain amplitude of the generated QFTA wave to that of the QLA wave are shown in Fig. 2(a), and Fig. 2(b), respectively. The single acoustic mode measured in the (100) sample is associated with coherent acoustic phonons with pure longitudinal (LA) polarization. Along the (311) orientation, denoted in the inverse pole figures in Fig. 2 with the white marker, we see that the ratio QSTA/QLA is ~0.025, while the ratio QFTA/QLA is ~0.15. Note that the inverse pole plots in Fig. 2 represent the relative excitation amplitude of the shear strain displacement for all possible crystallite orientations, and do not consider detection efficiencies of the shear modes. Transverse acoustic modes are generated along the (311) direction due to mode conversion at the gold film/ thoria substrate interface [51, 55]. While the amplitude of the QSTA mode in the (311) sample is very low, and within the detection noise floor, we attribute the dominant detected peak in the (311) sample to being the QLA mode, while the second, lower amplitude peak is attributed to the QFTA mode (i.e., a mode with predominantly shear polarization), given that the amplitude ratio QFTA/QLA predicted by the model is in reasonable agreement with that observed in the experiment.

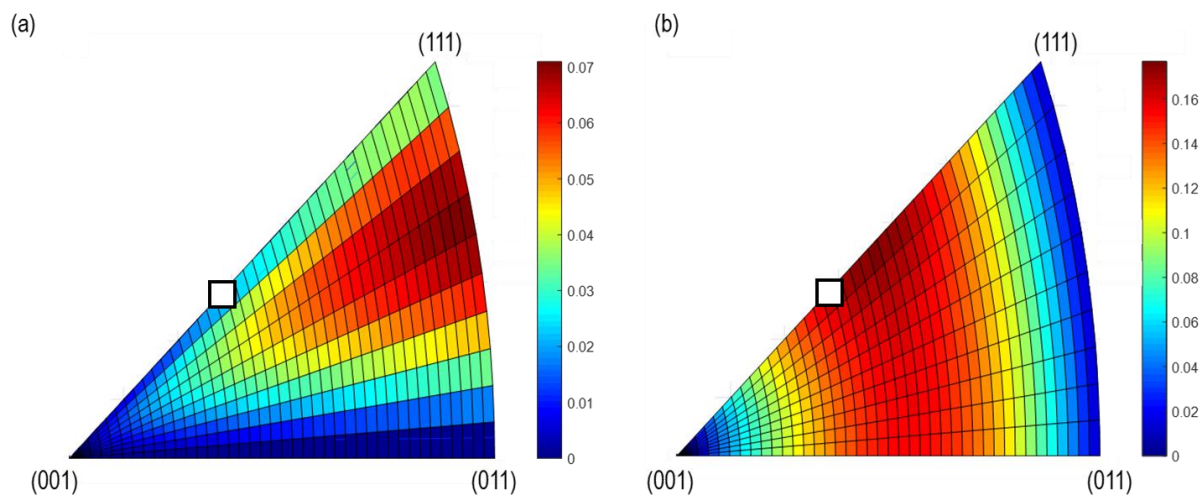

Figure 2. Color-coded inverse pole figures showing the calculated displacement amplitude ratios of (a) quasi-slow transverse acoustic (QSTA) mode to quasi-longitudinal acoustic (QLA) mode and (b) quasi-fast transverse acoustic (QFTA) to quasi-longitudinal acoustic (QLA) mode along different crystallite orientations in thoria. The white square highlights the displacement ratios (QSTA/QLA QFTA/QLA) along the (311) direction.

The intensity of the detected signal is governed by the amplitude of generated strain pulse as well as the photoelastic tensor of the crystal [55]. Following the methodology presented in reference [55], we measured the amplitude of the Fourier peak of the acoustic mode as a function of the optical polarization angle of the probe beam (defined as the angle between the optical polarization vector of the probe beam and the $x_1$ axis of the sample, shown in Fig. 1(a)). Figures 3(a) and (b) show the influence of the amplitude of detected LA mode along the (100) orientation, and that of the QLA and QFTA modes along the (311) orientation, respectively. The amplitude of the Fourier peak associated with the LA mode along the (100) orientation is largely insensitive to the probe polarization angle. The dashed blue line in Fig. 3(a), shown as a guide to the eye, captures the observed trend. The amplitudes of the QLA and QFTA modes along the (311) orientation, on the other hand, exhibit a distinct dependence on the probe polarization angle and exhibit a two-fold rotational symmetry. These trends are similar to profiles reported previously for ceria crystallites [55]

The amplitudes of the QFTA modes have been normalized with respect to the maximum measured amplitude of the QLA mode. The dependence of the detected amplitudes of the QLA and QFTA modes on the probe polarization angle is due to anisotropy of the photoelastic tensor of thoria [55]. The solid curves in Fig. 5(b) represent model profiles obtained by optimizing the photoelastic tensor ratio and the best fit is obtained when the ratio of the photoelastic tensor terms $(P_{11} - P_{12} - 2P_{44})/P_{12}$ = -1.2. The above procedure allowed us to optimize the optical polarization angle of the probe beam to maximize the intensity of the shear mode, a critical step for calculating the elastic stiffness tensor components.

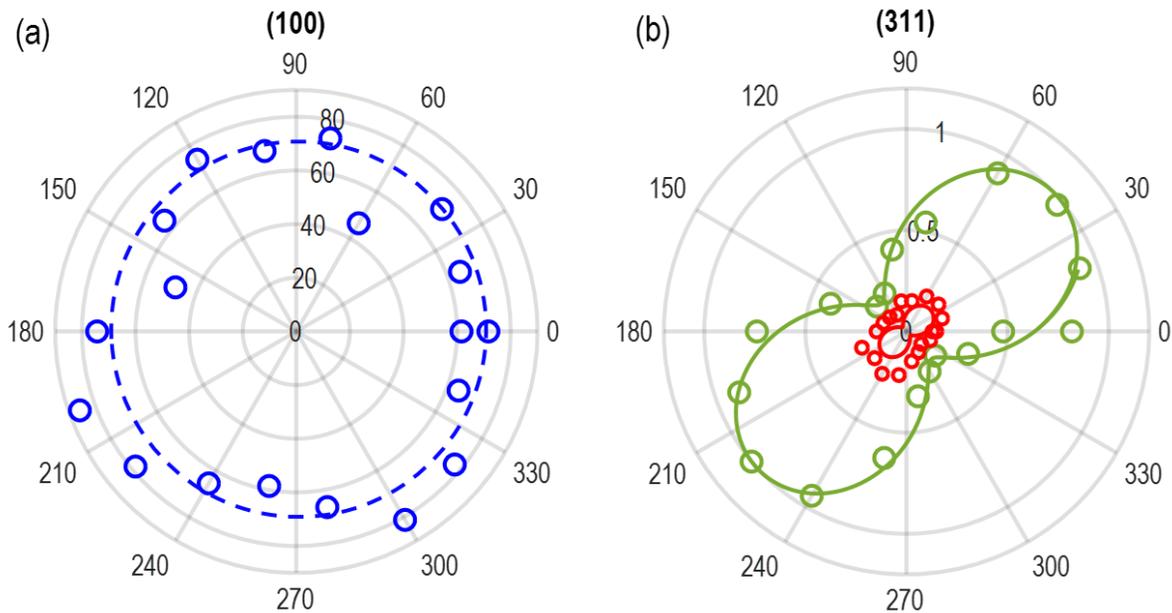

Figure 3. Polar plots of the (a) LA mode in (100) (blue circle markers) and (b) QLA mode (green circle markers) and normalized QFTA mode (red circle markers) in (311) thoria as a function of probe polarization angle. The dashed curve in panel (a) is a guide to the eye, while the solid curves in panel (b) represent the theoretical amplitude profiles from the optimization of the photoelastic tensor ratio.

*3.4. Temperature-dependent elastic constants*

Next, we tracked the change in the Brillouin oscillation frequencies of the LA mode in the (100) sample, and the QLA and QFTA modes in the (311) crystals over 77 - 350 K temperature range. Temperature-dependent TDBS signals on the (311) surface-oriented crystal were acquired by setting the relative polarization angle of the probe to maximize the amplitude of the detected QFTA mode. Figures 4(a) – (c) show the Fourier spectra for each of the three modes at 77 K (blue curve) and at 350 K (red curve). In all cases, we see a decrease in frequency, or 'softening' of the acoustic mode with temperature. The frequency peaks were identified by fitting a Gaussian function to the Fourier peak. In most cases, the full width at half-maximum of the measured acoustic modes was < 0.9 GHz. We note that the width of the peak is limited by the finite number of Brillouin oscillations within the measured ~1.5 ns duration time trace. Decay of the Brillouin oscillations can be attributed to several factors including the coherence time of the laser beams, Rayleigh range of the probe beam, and decay due to optical and ultrasonic attenuation in the medium [56]. These factors also contribute to the observed finite Fourier linewidth of the detected coherent acoustic modes.

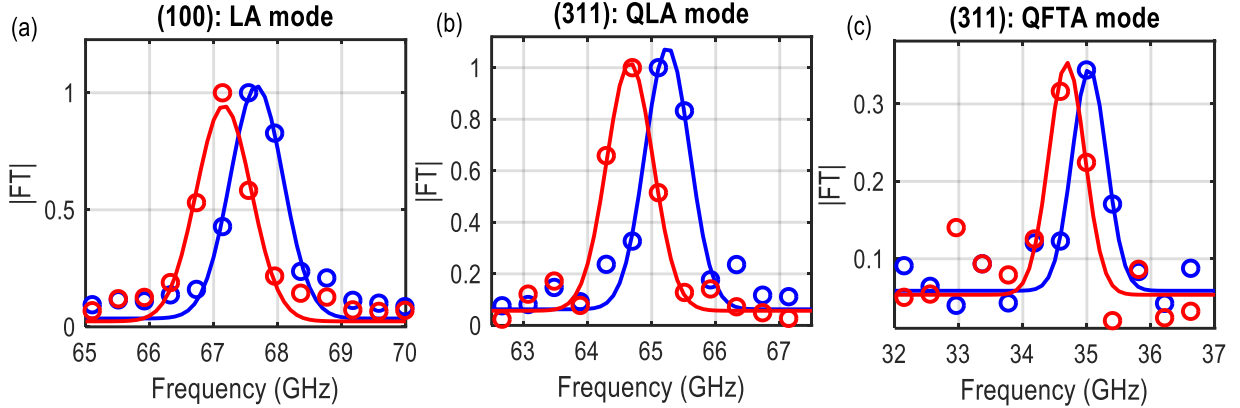

Figure 4. Power spectra highlighting the frequency peaks of the (a) LA mode in the (100) crystal; (b) QLA mode; and (c) QFTA mode in the (311) crystal associated with Brillouin oscillations measured at 77 K (blue curve) and 350 K (red curve). The solid lines denote the Gaussian fits to the measured frequency peaks.

While the temperature-dependent refractive index of thoria was not measured, we expect this change to be within the measurement uncertainty, since the bandgap of thoria (5.4 eV) and the critical points in the dielectric function are in the extreme ultraviolet, considerably far from the optical probing wavelength (390 nm) [57]. A previous study [58] on the temperature-dependent refractive index of calcium fluoride ($CaF_2$), a fluorite crystal similar in structure to thoria and also with a wide bandgap [59], showed that the index of refraction of $CaF_2$ reduced by only 0.12% from 70 K to 295 K at 400 nm. From the Eq. (iii), we see that the elastic constants $C_{ij}$ are proportional to $\rho v^2$. The temperature-dependence of the density of thoria was used when calculating $C_{ij}$ and was calculated from previously reported temperature-dependent values of the thermal expansion coefficient of thoria [60]. The calculated density of thoria reduced from 10.03 g/cm$^3$ at 77 K to 9.98 g/cm$^3$ at 350 K.

Figure 5 illustrates the temperature-dependence of the three elastic constants ($C_{11}$, $C_{12}$, and $C_{44}$). $C_{11}$ was calculated directly from the measured Brillouin oscillation frequency of the LA mode in the (100) sample, using the relation, $C_{11} = \rho v_{LA}^2$. While $C_{11}$ was calculated directly from a single measurement of the (100) sample, $C_{12}$ and $C_{44}$ were calculated from the ultrasonic velocities of the QLA and QFTA modes in the (311) samples using the inverse problem of the Christoffel equation (i.e., solving for the elastic constants from the measured ultrasonic wave velocities along (311)), and using the value of $C_{11}$ obtained from the Brillouin frequency measured along the (100) orientation). We note that while the two acoustic modes measured along the (311) orientation also have sensitivity to $C_{11}$, the frequencies of these modes were used to only calculate $C_{12}$ and $C_{44}$. From Fig. 4(a), we see that $C_{11}$ monotonically decreases with temperature above 100 K, and appears to start to asymptote to a value of ~367 GPa below 100 K. This follows the general, qualitative trend of the change in the change in elastic stiffness with temperature, as described by Ledbetter [47]: monotonic decrease with increasing temperature, and approaching linearity at higher temperatures, and flattening out (i.e., $dC/dT = 0$) at zero temperature. $C_{44}$ shows a similar trend, albeit with a larger scatter. $C_{12}$ measurements also show a large scatter, making it hard to discern any clear trend with temperature. The error bars in Fig. 4 represent the uncertainties in the calculated elastic constants and were calculated using a Monte-Carlo routine by taking the uncertainties in the measured Brillouin oscillation frequency as inputs. Trends in the temperature dependence of the elastic stiffness tensor components have been investigated in other materials with fluorite crystal structure, such as in calcium fluoride ($CaF_2$) [61] and urania [62] where a similar trend was reported, although additional temperature-dependent structure around 30 K was observed in urania due to the paramagnetic-to-antiferromagnetic phase transition [62].

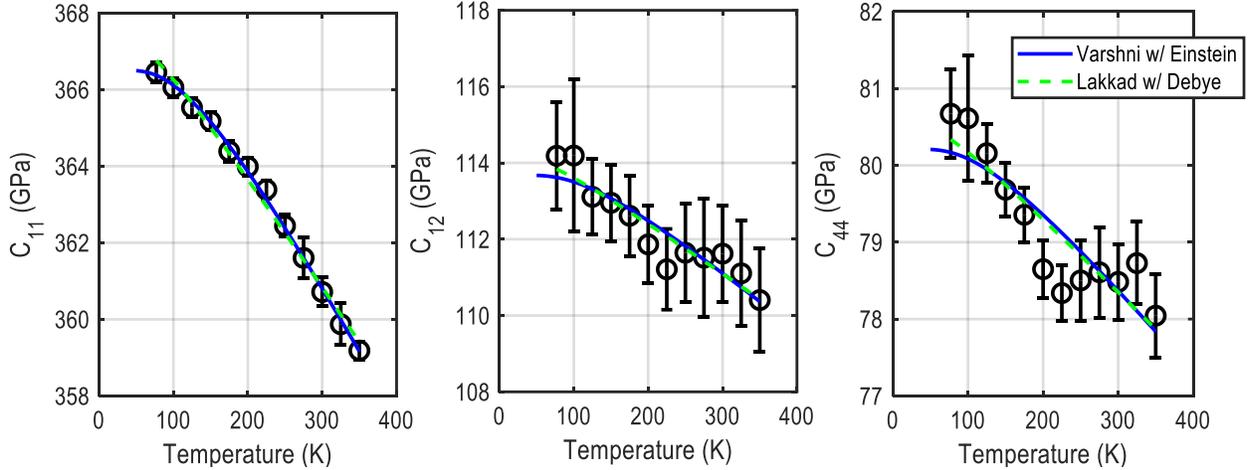

Figure 5. Temperature-dependence of the elastic constants calculated from the measured Brillouin frequencies. The black error bars are associated with the uncertainty in the Brillouin frequency by fitting a Lorentzian function to the Brillouin frequency peaks. The solid blue line and the dashed green lines denote fits to the Varshni/Einstein and Lakkad/Debye models using the calculated Debye temperature from the elastic constants at 77 K.

## IV. DISCUSSION

### 4.1. Analysis of anharmonic properties

Despite the larger scatter in the change of $C_{12}$ and $C_{44}$ with temperature, we fit the measured temperature-dependence of the three elastic constants to established semi-empirical relations proposed by Varshni [36] and Lakkad [37], that are based on the Leibfried-Ludwig relation in Eq. (i). While these semi-empirical models lack rigorous theoretical treatment of the phonon structure [63], their simplified treatment of phonon structure, and description of harmonic and anharmonic properties using a set of three parameters have been shown to adequately capture the temperature-dependence of the elastic constants measured in a variety of metal and semiconductor single crystals [22, 28, 34, 47]. Varshni used the energy of an Einstein oscillator to obtain the following elastic stiffness-temperature relation:

$$C_V(T) = C_{0,V} - \frac{s}{e^{(T_E/T)} - 1} \qquad \text{(iii)}$$

with $C_{0,V}$, $s$ and $T_E$ being adjustable parameters. $C_{0,V}$ represents the zero-temperature elastic stiffness. $s = 2(C^* - C_0)$ represents the deviation from the elastic stiffness $C^*$ obtained from linear extrapolation from high temperatures [47]. $T_E$ is a parameter that is related to the Einstein temperature $\theta_E$ of the crystal. Lakkad provided another simplified relation using a phenomenological model that consisted of a harmonic oscillator with the applied force term plus a third-order term representing the anharmonicity of the oscillator, obtained from the Debye model [37]. Lakkad's relation is of the form:

$$C_L(T) = C_{0,L}\left[1 - KF\left(T/T_D\right)\right] \tag{iv}$$

where

$$F\left(T/T_D\right) = 3\left(T/T_D\right)^4 \int_0^{T/T_D} [x^3\{e^x - 1\}^{-1}]dx \tag{v}$$

Here, $C_{0,L}$ represents the elastic stiffness at zero-temperature, $K$ represents the effect of zero-point vibration energy and captures the anharmonic properties of the crystal, and $T_D$ is related to the Debye temperature of the solid.

We determined that for reliable fitting of our results using the above models, it was best to use $T_D$ and $T_E$ as known and use two fitting parameters in the Varshni and Lakkad functions ($C_0$, $s$ and $E_0$, $K$, respectively). Next, we describe the procedure for evaluating these quantities using our results. We estimate the Debye temperature $\theta_D$ of thoria from the elastic constants measured at 77 K using the relation [64]:

$$\theta_D = \frac{h}{k_B}\left[\frac{3n_M}{4\pi}\left(\frac{N_A\rho}{M}\right)\right]^{1/3} v_m, \tag{vi}$$

where $h$ and $k_B$ are the Planck's constant and the Boltzmann constant, respectively. $n_M$, $N_A$ and $M$ denote the number of atoms in a thoria molecule (i.e., $n_M = 3$), Avogadro's number and the molecular weight of thoria, respectively. $v_m = \left[\frac{1}{3}\left(\frac{2}{v_s^3} + \frac{1}{v_l^3}\right)\right]^{-1/3}$ represents an average wave velocity integrated over several crystallographic directions [64]. We calculate the isotropic polycrystalline longitudinal and shear wave velocities using $v_l = \left(\frac{B+\frac{4}{3}G}{\rho}\right)^{1/2}$ and $v_s = \left(\frac{G}{\rho}\right)^{1/2}$, where $B = (C_{11} + 2C_{12})/3$ is the bulk modulus and $G_V = (C_{11} - C_{12} + 3C_{44})/5$ and $G_R = \frac{5(C_{11}-C_{12})C_{44}}{4C_{44}+3(C_{11}-C_{12})}$ are the Voigt and Reuss shear moduli, respectively [65]. Using 77 K elastic constants, we calculate the Debye temperature, $\theta_D = 427$ K using the Voigt moduli, and $\theta_D = 417$ K using the Reuss moduli. These values are in reasonable agreement with values determined using DFT ($\theta_D = 402.6$ K) [66], while lower values have been reported from heat capacity measurements ($\theta_D = 259$ K) [67] [68], room-temperature elastic constants measurements ($\theta_D = 290$ K) [39], and from neutron diffraction data at several temperatures ($\theta_D = 268$ K) [69].

In simple cases, the Einstein temperature can be calculated from the Debye temperature, $\theta_E = 0.75\theta_D = 351$ K [34, 47, 70]. We use $T_E = \theta_E = 351$ K as in input parameter in Varshni function, equation (iii), and $\theta_D = 468$ K as an input parameter in the Lakkad function, equations (iv) and (v), to fit to our measurements. The solid blue line and the dashed green lines in Fig. 4 denote fits to the Varshni and the Lakkad models, respectively. The parameters obtained from the fit (along with 95% confidence intervals) are listed in Table 1. We see that both models capture the general trend of elastic 'softening' with increasing temperature in all cases. The narrow bounds of the 95% confidence interval on the parameters fitted to the $C_{11}$ data confirm clear monotonic decrease in the absolute value of $C_{11}$ over the measured temperature range. This trend is indicative of acoustic mode softening duet to lattice anharmonicity. As would be expected from the observed scatter, the 95% confidence interval bounds on the parameters obtained by fitting to the $C_{12}$ and $C_{44}$ are slightly larger than those obtained by fitting $C_{11}$.

Table 1. Values of parameters (along with 95% confidence intervals) obtained by fitting the measured temperature-dependent elastic constants to the Varshni and Lakkad relations.

| $C_{ij}$ | Varshni function (with Einstein model) | | Lakkad relation (with Debye model) | |
|---|---|---|---|---|
| | $C_{0,V}$ (GPa) | $s$ | $C_{0,L}$ (GPa) | $K$ |
| $C_{11}$ | $366.51 \pm 0.04$ | $12.68 \pm 0.45$ | $367.44 \pm 0.34$ | $0.0177 \pm 0.0012$ |
| $C_{12}$ | $113.68 \pm 0.56$ | $5.72 \pm 1.77$ | $114.15 \pm 0.59$ | $0.026 \pm 0.007$ |
| $C_{44}$ | $80.21 \pm 0.53$ | $4.10 \pm 1.67$ | $80.56 \pm 0.57$ | $0.027 \pm 0.009$ |

## 4.2. Uncertainty in $C_{12}$ and $C_{44}$ from TDBS measurements

To investigate the reasons behind the scatter in the measured temperature-dependent trends in $C_{12}$ and $C_{44}$, we use Christoffel's equation to analyze the theoretical sensitivity of the elastic constants calculated from a known acoustic velocity along a prescribed crystallographic orientation. $C_{11}$ is sensitive only to changes in the LA mode velocity along the (100) direction. Since $C_{11} = v_{LA}^2 \rho$, it is clear that a 1% uncertainty in $v_{LA}$ results in a 2% uncertainty in $C_{11}$. On the other hand, uncertainties of 1% in the QLA and QFTA mode velocities along the (311) orientation yield corresponding uncertainties of 14% in $C_{12}$ and 8% in $C_{44}$. The limited theoretical sensitivity to $C_{12}$ and $C_{44}$, coupled with the lower amplitude of the generated QFTA mode, can therefore explain the observed scatter in the measured temperature-dependence of $C_{12}$ and $C_{44}$. TDBS measurements in thoria crystals along other lower-symmetry orientations that generate all three modes (QLA, QFTA and QSTA) with sufficiently high excitation amplitude and detection efficiency may help in greater precision in the temperature-dependence of the elastic constants. RUS is another versatile laboratory technique used for measuring second-order elastic constants of solids from the mechanical resonances of samples with well-defined geometry [71]. Unlike TDBS, RUS does not require multiple measurements on samples with differing crystallographic orientations [71]. Thus, while RUS can provide measurements of elastic constants with high accuracy, certain limitations associated with this technique, such as the high sensitivity to errors in geometry and homogeneity of the sample, sub-optimal boundary and coupling conditions, and difficulties in measuring crystals with irregular geometries, can be overcome by TDBS.

## 4.3. Comparison with elastic constants reported in the literature

We also compare the elastic constants of thoria measured using TDBS to previous experimentally obtained and theoretically calculated values, as shown in Table 2. We find excellent agreement with the room temperature elastic constants measured using the ultrasonic pulse-echo technique [39]. The elastic constants measured using INS in reference [40] are consistently higher than our measured values. This discrepancy could result from the inaccuracy of parametrization of rigid ion [72, 73] and core shell [74] models in fitting the INS measured phonon dispersion and the resulting propagation of the uncertainty to calculation of elastic constants. The room temperature value of $C_{44}$, extracted from the transverse acoustic mode along the Γ to $X_z$ direction in the time-of-flight INS experiment [21] shows excellent agreement with the corresponding value reported by Macedo et al. from pulse-echo measurements [39]. Table 2 also lists the ground state (0 K) elastic constants of thoria obtained from DFT calculations. Considerable scatter is observed in the elastic constants obtained using various approximation functionals in DFT. The elastic constants at 300 K reported by Cooper et al. using molecular dynamics (MD) simulations are also listed in Table 2. We also compare the temperature dependence of the elastic constants obtained from our TDBS measurements with corresponding calculations recently reported using DFT with three exchange-correlation functionals [21]. We note that in our experiment, since the measured Brillouin oscillation frequencies are in the several tens of gigahertz, ultrasonic wave generation in the sample can be considered an adiabatic process, and therefore provide a direct comparison to the calculated adiabatic elastic constants. Figure 6(a) – (c) illustrates the comparison. DFT calculations using the SCAN functional provide the closest agreement to our experimental values. Figure 6(d) shows the temperature dependence of the bulk modulus B calculated using modulus $B = \frac{1}{3}(C_{11} + 2C_{12})$ from the measured values of $C_{11}$ and $C_{12}$. We find that within the uncertainties, there is reasonable agreement with the temperature-dependence predicted by the DFT [11, 17]. The MD study by Cooper et al. [8] are found to underpredict the bulk modulus obtained from TDBS measurements at 300 K. It is interesting to note that while the two DFT studies [11, 17] predict bulk moduli with reasonable agreement at low temperatures, they show considerable differences above 350 K.

Table 2. Comparison of the room temperature elastic constants obtained from the TDBS measurements with other experimental and computational studies.

|  | $C_{11}$ (GPa) | $C_{12}$ (GPa) | $C_{44}$ (GPa) |
|---|---|---|---|
| TDBS experiment (this work) | 360.71 ± 0.375 | 111.62 ± 1.272 | 79.48 ± 0.492 |
| *Experimental studies*[§] | | | |
| Macedo et al. (Pulse-echo) [39] | 367 ± 4 | 106 ± 2 | 79.7 ± 0.8 |
| Clausen et al. (INS) [40] | 377 | 146 | 89 |
| Mathis et al. (Time-of-flight INS) [21] | - | - | 80.77 |
| *Computational studies* | | | |
| Szpunar et al. (DFT GGA – PBE) [17][¶] | 351.9 ± 2.1 | 105.4 ± 1.4 | 70.9 ± 0.4 |
| Szpunar et al. (DFT GGA – PBEsol) [17][¶] | 370.9 ± 1.9 | 118.7 ± 1.4 | 80.8 ± 0.3 |
| Szpunar et al. (DFT GGA – WC) [17][¶] | 370.6 ± 1.2 | 119.3 ± 0.8 | 370.6 ± 1.2 |
| Cooper et al. (Mol. Dynamics – LAMMPS) [8][†] | 352.3 | 113.4 | 71.7 |
| Lu et al. (DFT GGA) [11][¶] | 351.2 | 106.9 | 74.1 |
| Lu et al. (DFT with spin-orbit coupling) [11][¶] | 348.4 | 107.3 | 72.1 |
| Terki et al. (DFT GGA) [19][¶] | 355 | 106 | 54 |
| Shein et al. (DFT FLAPW) [15][¶] | 314.5 | 73.1 | 75.7 |
| Mathis et al. (DFT LDA) [21][¶] | 376.5 | 127.5 | 85.1 |
| Mathis et al. (DFT GGA) [21][¶] | 345.3 | 106.3 | 70.3 |
| Mathis et al. (DFT SCAN) [21][¶] | 367.3 | 114.4 | 79.9 |

[§] Room temperature (300 K) measurements
[¶] Ground state (0 K) property calculation
[†] Room temperature (300 K) property calculation

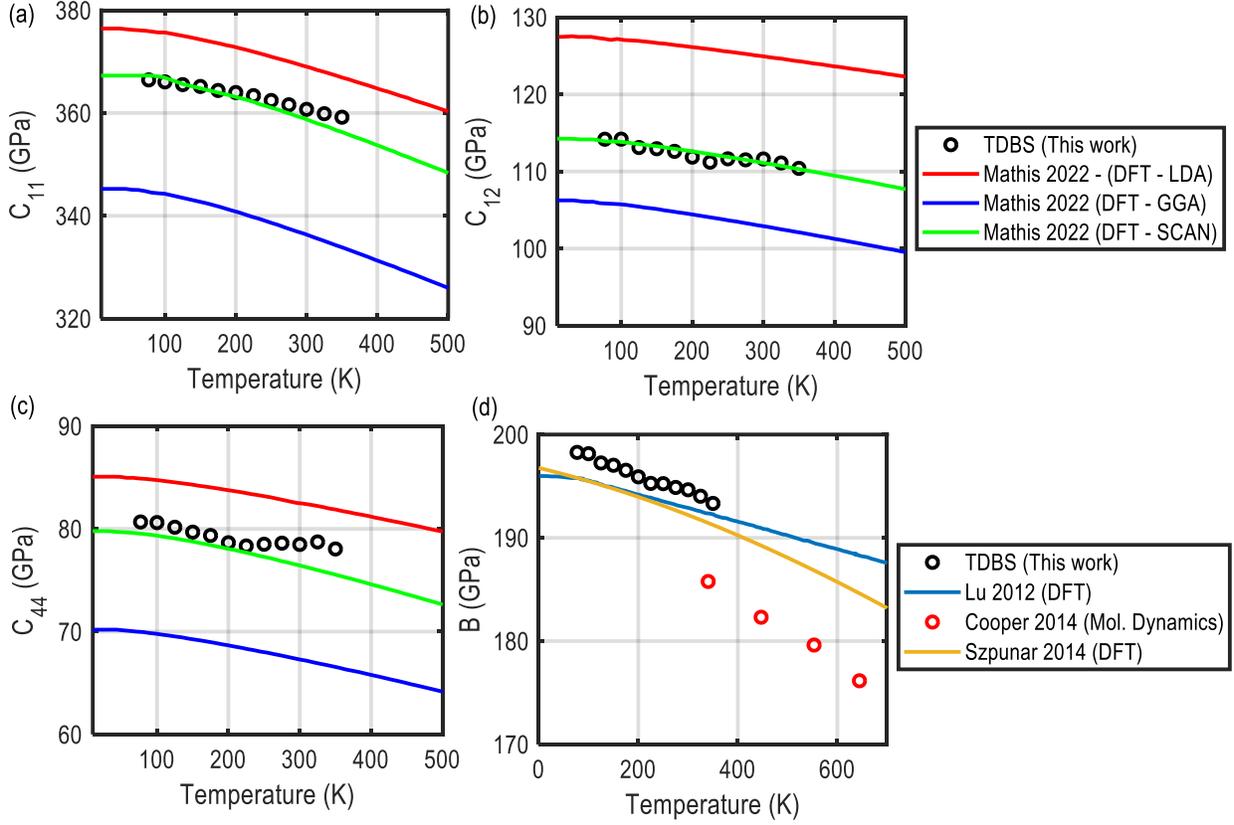

Figure 6. (a) – (c) Comparison of the measured temperature dependence of the elastic constants C11, C12 and C44, respectively, with recently reported theoretical calculations [21]. (d) The experimental bulk modulus as a function of temperature compared to two theoretical works [8, 11, 17].

*4.4. Temperature-dependent Gruneisen parameter obtained from measured elastic constants*

Our measurements allow us to evaluate another common measure of lattice anharmonicity, namely thermodynamic Gruneisen parameter $\gamma$ [75, 76]. While the mode Gruneisen parameter describes the phonon frequency shift with respect to the volume for a particular acoustic mode, the acoustic Gruneisen parameter represents the weighted average of the mode Gruneisen parameter for all acoustic phonon branches [11]. The acoustic Gruneisen parameter can also be defined as [29, 77],

$$\gamma(T) = \frac{\alpha_V(T) B(T)}{C_p(T) \rho(T)} \quad \text{(vi)}$$

where $\alpha_V(T)$, $B(T)$, $C_p(T)$ and $\rho(T)$ are the temperature-dependent volume thermal expansion coefficient, bulk modulus, specific heat capacity at constant pressure and density, respectively. We calculate the acoustic Gruneisen parameter from the bulk moduli measured at various temperatures, along with the temperature-dependent density obtained from previously reported temperature dependent thermal expansion coefficient [60], and recently experimentally measured heat capacity $C_p(T)$ [78]. The temperature dependent Gruneisen parameter is shown in Fig. 7. The acoustic Gruneisen parameter varies from 2.6 to 2.15 over 77 – 350 K. In comparison, Sobolev and Lemehov assumed a constant (i.e., temperature independent) Gruneisen parameter $\gamma = 1.8$ to theoretically calculate the heat capacity, thermal expansion coefficient and the thermal conductivity of thoria [16].

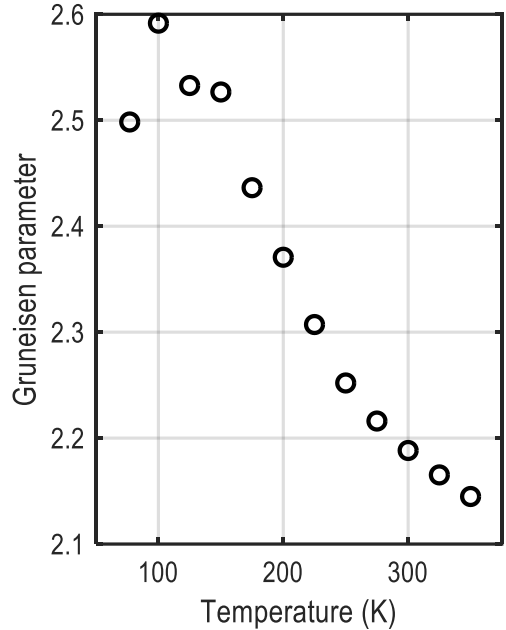

Figure 7. Thermodynamic Gruneisen parameter calculated from the measured temperature-dependent bulk modulus and previously reported values of volume thermal expansion coefficient, density, and heat capacity at constant pressure.

Lastly, we'd like to emphasize that the demonstrated capability of measuring temperature dependent elastic constants on a small-scale sample and quantifying lattice anharmonicity has important implications for validation of first principles calculations of materials properties. The current method provides an indirect measurement of both harmonic and anharmonic force constants, all important for evaluation and prediction of temperature dependence of elastic constants [21], lattice specific heat, thermal expansion, and thermal conductivity [75]. Especially, presented experimental method provide necessary measurement for either validation of *ab initio* calculation or training empirical interatomic potentials [79, 80]. The latter can be the only atomistic approach in a system where reliable DFT calculations are not trivial as in the case of strongly electron correlated systems [81].

V. CONCLUSIONS

Time-domain Brillouin scattering (TDBS) was used to measure the temperature-dependence of the adiabatic elastic constants of single crystal thorium dioxide from 77 K – 350 K. The elastic constants show a ~2.5% reduction over the measured temperature range arising from lattice anharmonicity. Given the rapidly growing interest in the use of thorium dioxide as an alternative fuel to uranium dioxide in fissile nuclear reactors, our measurements of temperature-dependent elastic constants influenced by lattice anharmonicity provide a means of validating theoretical models aimed at predicting vital thermophysical properties such as the thermal conductivity, thermal expansion coefficient and heat capacity. While our measurements reveal a smooth and monotonic reduction of $C_{11}$ with temperature, considerable scatter was observed in the the measured temperature-dependence of $C_{12}$ and $C_{44}$. This scatter is attributed to the lower detection sensitivity of the fast transverse mode along the low-symmetry (311) crystallographic direction. By analyzing the dependence of the amplitude of the quasi-transverse and quasi-longitudinal modes on the optical polarization angle of the probe laser beam, we determine the photoelastic tensor anisotropy ratio in thoria. Measurement of the photoelastic anisotropy ratio can be used to determine specific of crystal orientations to maximize the theoretical sensitivity of the elastic constants to the ultrasonic velocities measured using TDBS, in addition to maximizing the generation and detection sensitivities. Such measurements could provide higher-fidelity temperature-dependent measurements of the complete elastic stiffness tensor components of thoria. Finally, we have calculated the temperature-dependent Gruneisen parameter using our temperature-dependent measurements of elastic constants and previously reported temperature-dependent measurements of heat capacity, thermal expansion coefficient and density. Future work investigating pressure-dependent measurements using the experimental methodology presented here will provide comprehensive benchmarks for advanced *ab initio* theoretical model predictions.

## VI. ACKNOWLEDGEMENTS

This work was supported by the Center for Thermal Energy Transport under Irradiation (TETI), an Energy Frontier Research Center funded by the US Department of Energy, Office of Science, Office of Basic Energy Sciences.

## VII. DATA AVAILABILITY

The data that support the findings of this study are available from the corresponding author upon reasonable request.

## VIII. REFERENCES:

1. Leibfried, G. and W. Ludwig, *Theory of anharmonic effects in crystals*, in *Solid state physics*. 1961, Elsevier. p. 275-444.
2. Barron, T., *Interatomic potentials in ideal anharmonic crystals.* Discussions of the Faraday Society, 1965. **40**: p. 69-75.
3. Cowley, R.A., *Anharmonic crystals.* Reports on Progress in Physics, 1968. **31**(1): p. 123.
4. Belle, J. and R. Berman, *Thorium dioxide: properties and nuclear applications*. 1984, USDOE Assistant Secretary for Nuclear Energy.
5. Bharadwaj, S., *Thoria-based nuclear fuels*. 1 ed. Green Energy and Technology, ed. S.R.B. Dasarathi Das. 2013, London: Springer.
6. Hurley, D.H., et al., *Thermal energy transport in oxide nuclear fuel.* Chemical Reviews, 2021. **122**(3): p. 3711-3762.
7. Behera, R.K. and C.S. Deo, *Atomistic models to investigate thorium dioxide (ThO2).* Journal of Physics: Condensed Matter, 2012. **24**(21): p. 215405.
8. Cooper, M., M. Rushton, and R. Grimes, *A many-body potential approach to modelling the thermomechanical properties of actinide oxides.* Journal of Physics: Condensed Matter, 2014. **26**(10): p. 105401.

9. Kanchana, V., et al., *First-principles study of elastic properties of CeO2, ThO2 and PoO2.* Journal of Physics: Condensed Matter, 2006. **18**(42): p. 9615.
10. Liu, J., et al., *Lattice thermodynamic behavior in nuclear fuel ThO2 from first principles.* Journal of Nuclear Materials, 2018. **511**: p. 11-17.
11. Lu, Y., Y. Yang, and P. Zhang, *Thermodynamic properties and structural stability of thorium dioxide.* Journal of Physics: Condensed Matter, 2012. **24**(22): p. 225801.
12. Ma, J.-J., et al., *Molecular dynamics study on thermal properties of ThO2 doped with U and Pu in high temperature range.* Journal of Alloys and Compounds, 2015. **627**: p. 476-482.
13. Nakamura, H. and M. Machida, *High-temperature properties of thorium dioxide: A first-principles molecular dynamics study.* Journal of Nuclear Materials, 2016. **478**: p. 56-60.
14. Park, J., E.B. Farfán, and C. Enriquez, *Thermal transport in thorium dioxide.* Nuclear Engineering and Technology, 2018. **50**(5): p. 731-737.
15. Shein, I., K. Shein, and A. Ivanovskii, *Elastic and electronic properties and stability of SrThO3, SrZrO3 and ThO2 from first principles.* Journal of nuclear materials, 2007. **361**(1): p. 69-77.
16. Sobolev, V. and S. Lemehov, *Modelling heat capacity, thermal expansion, and thermal conductivity of dioxide components of inert matrix fuel.* Journal of nuclear materials, 2006. **352**(1-3): p. 300-308.
17. Szpunar, B. and J. Szpunar, *Theoretical investigation of structural and thermo-mechanical properties of thoria up to 3300 K temperature.* Solid state sciences, 2014. **36**: p. 35-40.
18. Szpunar, B., J. Szpunar, and K.-S. Sim, *Theoretical investigation of structural and thermo-mechanical properties of thoria.* Journal of Physics and Chemistry of Solids, 2016. **90**: p. 114-120.
19. Terki, R., et al., *First principles calculations of structural, elastic and electronic properties of XO2 (X= Zr, Hf and Th) in fluorite phase.* Computational materials science, 2005. **33**(1-3): p. 44-52.
20. Wang, B.-T., et al., *First-principles study of ground-state properties and high pressure behavior of ThO2.* Journal of Nuclear Materials, 2010. **399**(2-3): p. 181-188.
21. Mathis, M.A., et al., *The generalized quasiharmonic approximation via space group irreducible derivatives.* Physical Review B, 2022. **106**(014314): p. 23.
22. Cottam, R.t. and G. Saunders, *The elastic constants of GaAs from 2 K to 320 K.* Journal of Physics C: Solid State Physics, 1973. **6**(13): p. 2105.
23. Boyle, W. and R. Sladek, *Elastic constants and lattice anharmonicity of GaSb and GaP from ultrasonic-velocity measurements between 4.2 and 300 K.* Physical Review B, 1975. **11**(8): p. 2933.
24. Diederich, M.E. and J. Trivisonno, *Temperature dependence of the elastic constants of sodium.* Journal of Physics and Chemistry of Solids, 1966. **27**(4): p. 637-642.
25. Gutman, E. and J. Trivisonno, *Temperature dependence of the elastic constants of rubidium.* Journal of Physics and Chemistry of Solids, 1967. **28**(5): p. 805-809.
26. Smith, J. and J. Gjevre, *Elastic constants of yttrium single crystals in the temperature range 4.2–400 K.* Journal of Applied Physics, 1960. **31**(4): p. 645-647.
27. Pederson, D. and J. Brewer, *Elastic constants of cadmium fluoride from 4.2 to 295 K.* Physical Review B, 1977. **16**(10): p. 4546.

28. Kim, S. and H. Ledbetter, *Low-temperature elastic coefficients of polycrystalline indium.* Materials Science and Engineering: A, 1998. **252**(1): p. 139-143.
29. Adams, J.J., et al., *Elastic constants of monocrystal iron from 3 to 500 K.* Journal of applied physics, 2006. **100**(11): p. 113530.
30. Chu, F., et al., *Elastic properties of C40 transition metal disilicides.* Acta materialia, 1996. **44**(8): p. 3035-3048.
31. Haglund, A., et al., *Polycrystalline elastic moduli of a high-entropy alloy at cryogenic temperatures.* Intermetallics, 2015. **58**: p. 62-64.
32. He, Y., et al., *Elastic constants and thermal expansion of single crystal γ-TiAl from 300 to 750 K.* Materials Science and Engineering: A, 1997. **239**: p. 157-163.
33. Isaak, D., et al., *Elasticity of TiO2 rutile to 1800 K.* Physics and Chemistry of Minerals, 1998. **26**(1): p. 31-43.
34. Sonehara, T., et al., *Temperature dependence of the Brillouin frequency shift in crystals.* Journal of applied physics, 2007. **101**(10): p. 103507.
35. Kang, D.H., et al., *Elastic Properties of Taurine Single Crystals Studied by Brillouin Spectroscopy.* International journal of molecular sciences, 2021. **22**(13): p. 7116.
36. Varshni, Y., *Temperature dependence of the elastic constants.* Physical Review B, 1970. **2**(10): p. 3952.
37. Lakkad, S.C., *Temperature dependence of the elastic constants.* Journal of Applied Physics, 1971. **42**(11): p. 4277-4281.
38. Siethoff, H. and K. Ahlborn, *The dependence of the Debye temperature on the elastic constants.* physica status solidi (b), 1995. **190**(1): p. 179-191.
39. Macedo, P., W. Capps, and J. Wachtman Jr, *Elastic constants of single crystal ThO2 at 25° C.* Journal of the American Ceramic Society, 1964. **47**(12): p. 651-651.
40. Clausen, K., et al., *Inelastic neutron scattering investigation of the lattice dynamics of ThO 2 and CeO 2.* Journal of the Chemical Society, Faraday Transactions 2: Molecular and Chemical Physics, 1987. **83**(7): p. 1109-1112.
41. Wachtman Jr, J., et al., *Exponential temperature dependence of Young's modulus for several oxides.* Physical review, 1961. **122**(6): p. 1754.
42. Wolfe, R. and S. Kaufman, *Mechanical Properties of Oxide Fuels (LSBR/LWB Development Program)*. 1967, Bettis Atomic Power Lab., Pittsburgh, Pa.(US).
43. Phani, K.K. and D. Sanyal, *Elastic properties of porous polycrystalline thoria—A relook.* Journal of the European Ceramic Society, 2009. **29**(3): p. 385-390.
44. Spinner, S., L. Stone, and F. Knudsen, *Temperature dependence of the elastic constants of thoria specimens of varying porosity.* J. Res. Natl. Bur. Std., 1963. **67**(2): p. 93-100.
45. Gusev, V.E. and P. Ruello, *Advances in applications of time-domain Brillouin scattering for nanoscale imaging.* Applied Physics Reviews, 2018. **5**(3): p. 031101.
46. Yu, K., et al., *Brillouin oscillations from single Au nanoplate opto-acoustic transducers.* ACS nano, 2017. **11**(8): p. 8064-8071.
47. Ledbetter, H., *Sound velocities, elastic constants: Temperature dependence.* Materials Science and Engineering: A, 2006. **442**(1-2): p. 31-34.
48. Mann, M., et al., *Hydrothermal growth and thermal property characterization of ThO2 single crystals.* Crystal growth & design, 2010. **10**(5): p. 2146-2151.
49. Hurley, D., et al., *Coherent control of gigahertz surface acoustic and bulk phonons using ultrafast optical pulses.* Applied Physics Letters, 2008. **93**(11): p. 113101.

50. Wang, Y., et al., *Nondestructive characterization of polycrystalline 3D microstructure with time-domain Brillouin scattering.* Scripta Materialia, 2019. **166**: p. 34-38.
51. Wang, Y. and M. Khafizov, *Shear wave generation by mode conversion in picosecond ultrasonics: Impact of grain orientation and material properties.* Journal of the American Ceramic Society, 2021. **104**(6): p. 2788-2798.
52. Cahill, D.G., *Analysis of heat flow in layered structures for time-domain thermoreflectance.* Review of scientific instruments, 2004. **75**(12): p. 5119-5122.
53. Fujiwara, H., *Spectroscopic ellipsometry: principles and applications*. 2007: John Wiley & Sons.
54. Khafizov, M., et al., *Subsurface imaging of grain microstructure using picosecond ultrasonics.* Acta Materialia, 2016. **112**: p. 209-215.
55. Wang, Y., et al., *Imaging grain microstructure in a model ceramic energy material with optically generated coherent acoustic phonons.* Nature communications, 2020. **11**(1): p. 1-8.
56. Gusev, V.E., et al., *Theory of time-domain Brillouin scattering for probe light and acoustic beams propagating at an arbitrary relative angle: Application to acousto-optic interaction near material interfaces.* arXiv preprint arXiv:2107.05294, 2021.
57. Mock, A., et al., *Band-to-band transitions and critical points in the near-infrared to vacuum ultraviolet dielectric functions of single crystal urania and thoria.* Applied Physics Letters, 2019. **114**(21): p. 211901.
58. Leviton, D.B., B.J. Frey, and T.J. Madison. *Temperature-dependent refractive index of CaF2 and Infrasil 301*. in *Cryogenic Optical Systems and Instruments XII*. 2007. International Society for Optics and Photonics.
59. Heaton, R.A. and C.C. Lin, *Electronic energy-band structure of the calcium fluoride crystal.* Physical Review B, 1980. **22**(8): p. 3629.
60. Taylor, D., *Thermal expansion data.* Transactions and Journal of the British Ceramic Society, 1984. **83**(2): p. 32-37.
61. Singh, R., S. Mitra, and C. Rao, *Temperature variations of the elastic constants of CaF 2 and SrF 2 crystals.* Physical Review B, 1991. **44**(2): p. 838.
62. Brandt, O.G. and C.T. Walker, *Temperature Dependence of Elastic Constants and Thermal Expansion for U O 2.* Physical Review Letters, 1967. **18**(1): p. 11.
63. Garber, J. and A. Granato, *Theory of the temperature dependence of second-order elastic constants in cubic materials.* Physical Review B, 1975. **11**(10): p. 3990.
64. Anderson, O.L., *A simplified method for calculating the Debye temperature from elastic constants.* Journal of Physics and Chemistry of Solids, 1963. **24**(7): p. 909-917.
65. Chung, D. and W. Buessem, *The elastic anisotropy of crystals.* Journal of Applied Physics, 1967. **38**(5): p. 2010-2012.
66. Zhang, P., B.-T. Wang, and X.-G. Zhao, *Ground-state properties and high-pressure behavior of plutonium dioxide: Density functional theory calculations.* Physical Review B, 2010. **82**(14): p. 144110.
67. Ali, M. and P. Nagels, *Evaluation of the Debye temperature of thorium dioxide.* physica status solidi (b), 1967. **21**(1): p. 113-116.
68. Osborne, D.W. and E.F. Westrum Jr, *The heat capacity of thorium dioxide from 10 to 305 K. The heat capacity anomalies in uranium dioxide and neptunium dioxide.* The Journal of Chemical Physics, 1953. **21**(10): p. 1884-1887.

69. Willis, B., *Neutron diffraction studies of the actinide oxides I. Uranium dioxide and thorium dioxide at room temperature.* Proceedings of the Royal Society of London. Series A. Mathematical and Physical Sciences, 1963. **274**(1356): p. 122-133.
70. Girifalco, L.A., *Statistical physics of materials*. 1973: Wiley-Interscience.
71. Schwarz, R. and J. Vuorinen, *Resonant ultrasound spectroscopy: applications, current status and limitations.* Journal of Alloys and Compounds, 2000. **310**(1-2): p. 243-250.
72. Ganesan, S. and R. Srinivasan, *Lattice dynamics of calcium fluoride: Part I. Lyddane, sachs, teller formula, diffuse x-ray scattering, and specific heat.* Canadian Journal of Physics, 1962. **40**(1): p. 74-90.
73. Srinivasan, R., *Elastic constants of calcium fluoride.* Proceedings of the Physical Society, 1958. **72**(4): p. 566.
74. Elcombe, M. and A. Pryor, *The lattice dynamics of calcium fluoride.* Journal of Physics C: Solid State Physics, 1970. **3**(3): p. 492.
75. Togo, A. and I. Tanaka, *First principles phonon calculations in materials science.* Scripta Materialia, 2015. **108**: p. 1-5.
76. Ward, A., et al., *Ab initio theory of the lattice thermal conductivity in diamond.* Physical Review B, 2009. **80**(12): p. 125203.
77. Brugger, K. and T. Fritz, *Grüneisen gamma from elastic data.* Physical Review, 1967. **157**(3): p. 524.
78. Dennett, C.A., et al., *The influence of lattice defects, recombination, and clustering on thermal transport in single crystal thorium dioxide.* Apl Materials, 2020. **8**(11): p. 111103.
79. Miaomiao Jin and Marat Khafizov and Chao Jiang and Shuxiang Zhou and Chris Marianetti and Matthew Bryan and Michael, E.M.a.D.H.H., *Assessment of empirical interatomic potential to predict thermal conductivity in ThO2 and UO2.* Journal of Physics: Condensed Matter, 2021.
80. Zhou, S., et al., *Improving empirical interatomic potentials for predicting thermophysical properties by using an irreducible derivatives approach: The case of thorium dioxide.* arXiv preprint arXiv:2204.13685, 2022.
81. Zhou, S.X., et al., *Capturing the ground state of uranium dioxide from first principles: Crystal distortion, magnetic structure, and phonons.* Physical Review B, 2022. **106**(12).